\documentclass[12pt]{article}
\usepackage[dvipdfmx,hiresbb]{graphicx}
\usepackage{float}
\begin{document}
\global\arraycolsep=2pt 
\newcommand{\p}{{\bf p}}
\newcommand{\q}{{\bf q}}
\newcommand{\s}{{\bf s}}
\newcommand{\x}{{\bf x}}
\newcommand{\y}{{\bf y}}
\thispagestyle{empty} 
\begin{titlepage}    
\begin{flushright}
TNCT-1601
\end{flushright}
\vspace{1cm}
\topskip 3cm
\begin{center}
  \Large{\bf 
      Predicted value of  $0 \, \nu \beta \beta$-decay effective
       Majorana mass with error of lightest neutrino mass
     }  \\
\end{center}                                   
\vspace{0.6cm}
\begin{center}
Shinji Maedan            
  \footnote{ E-mail: maedan@tokyo-ct.ac.jp}   
           \\
\vspace{0.8cm}
  \textsl{ Department of Physics, Tokyo National College of Technology,
             Kunugida-machi,Hachioji-shi, Tokyo 193-0997, Japan}
  \end{center}                                               
\vspace{0.5cm}
\begin{abstract}
\noindent       
\\
Assuming that the lightest neutrino mass $ m_0 $ is measured, we study the
   influence of error of the measured $ m_0 $ on the uncertainty of the predicted
   value of the neutrinoless double beta decay ($0 \, \nu \beta \beta$)
   effective Majorana mass $  \vert m_{ee} \vert $.
The error of the predicted value of effective Majorana mass
   $ \sigma ( \vert m_{ee} \vert ) $ due to the error $ \sigma (m_0) $ is obtained
   by use of the propagation of errors.
We investigate in detail how $ \sigma ( \vert m_{ee} \vert ) $ behaves
   when the Majorana phases $ \beta $ and $ \alpha $ change, and also
   when the size of the lightest neutrino mass changes.
It is shown that $ \sigma ( \vert m_{ee} \vert ) $ does not exceed $ \sigma (m_0) $,
   and that the minimum value of $ \sigma ( \vert m_{ee} \vert ) $ in the
   $ \beta - \alpha $ plane can be zero if the size of $ m_0 $ is rather small.
In the normal mass ordering case, the maximum value of
   $ \sigma ( \vert m_{ee} \vert ) $  in the $ \beta - \alpha $ plane becomes about
   $ 0.68 \, \sigma (m_0) $ if $ m_0 (= m_1 ) $ is very small,
   while in the inverted mass ordering case, it  becomes about
   $ 0.02 \, \sigma (m_0) $ if $ m_0 (= m_3 ) $ is very small.
By use of the calculated $ \sigma ( \vert m_{ee} \vert ) $, we also discuss the
   minimum condition for the relative error of the lightest neutrino mass
   to raise the possibility of obtaining the information on $ \beta $ and $ \alpha $.
\\
\end{abstract} 
\end{titlepage}
%
%
\setcounter{page}{1}
\section{Introduction}
The experiments of neutrinoless double beta decay ($0 \, \nu \beta \beta$)
   play an important role in judging whether the neutrinos are Majorana particles
   or Dirac particles.
The neutrinoless double beta decay is the double beta decay which does not emit
   any neutrinos.
If we assume $0 \, \nu \beta \beta$ is caused by the exchange of three light Majorana
   neutrinos, the observation of  $0 \, \nu \beta \beta$ is the evidence that the neutrinos
   are Majorana fermions \cite{rf:DoiKotTak,rf:BilPet,rf:Rod}.
Many experiments of $0 \, \nu \beta \beta$ are in progress and planed:
   CANDLES  \cite{rf:UmeKisOga}, NEMO-3 \cite{rf:Arg}, SOLOTVINO \cite{rf:Dan},
   CUORICINO \cite{rf:Arn}, EXO-200 \cite{rf:Aug}, KamLAND-Zen \cite{rf:Gan}, etc.
The signal of $0 \, \nu \beta \beta$ has not been detected yet.
If neutrinos are Majorana fermions, the lepton mixing matrix $U$
   (MNS matrix \cite{rf:MakNakSak}) is expressed by six parameters;
   three lepton mixing angles $( \theta_{12}, \theta_{23}, \theta_{13})$, CP violating
   Dirac phase $\delta$, and two CP violating Majorana phases $\alpha, \beta$.
The two Majorana phases $\alpha, \beta$ are the degrees of freedom arising from
   the fact that the particle and the antiparticle are the same in Majorana fermion.
Among these six parameters, the value of three lepton mixing angles
   $( \theta_{12}, \theta_{23}, \theta_{13})$ are measured \cite{rf:Gon}.
Especially, the measurement of $ \theta_{13}$ was made recently \cite{rf:Abe},
   whose effect on the study of $0 \, \nu \beta \beta$ had been investigated \cite{rf:LinMerRod}.
The values of Dirac phase $\delta$ and two Majorana phases $\alpha, \beta$ are
   unknown to this day.
If we suppose that $0\, \nu \beta \beta$ is caused by the exchange of three light
   Majorana neutrinos, the amplitude of the decay is proportional to the 
   $0\, \nu \beta \beta$-decay effective Majorana mass $  \vert m_{ee} \vert $,
\begin{equation}
    \vert m_{ee} \vert
   =  \vert  \sum_{i=1}^{3} m_i \, U_{e i}^2  \vert
   =  \left\vert \, m_1  \vert U_{e 1} \vert^2 + m_2 \vert U_{e 2} \vert^2  e^{2 i \alpha}
    +  m_3 \vert U_{e 3} \vert^2  e^{2 i \beta} \,  \right\vert,
  \label{aa}
\end{equation}
where $m_i (i=1,2,3)$ is neutrino mass of $i$-th mass eigenstate.
The unitary matrix $U$ is the lepton mixing matrix (MNS matrix) and parametrized as follows,
\begin{eqnarray}
  U =  \left(   \begin{array}{ccc}
                 1  &        0       &   0           \\
                 0  &   c_{23}   &   s_{23}    \\
                 0  &   -s_{23}   &   c_{23}
                                       \end{array}
                                                     \right)
         \left(   \begin{array}{ccc}
                 c_{13}  &        0       &   s_{13} e^{-i \delta}         \\
                 0          &        1       &     0                                   \\
                 -s_{13} e^{ i \delta}          &   0   &   c_{13}
                                       \end{array}
                                                     \right)
         \left(   \begin{array}{ccc}
                  c_{12}   &   s_{12}   &   0       \\
                 -s_{12}   &   c_{12}   &   0       \\
                 0             &         0     &    1 
                                     \end{array}
                                                     \right)
          \left(   \begin{array}{ccc}
                       1      &      0                   &      0        \\
                      0       &     e^{ i \alpha}    &     0        \\
                       0      &      0                    &    e^{ i \beta}  
                                       \end{array}
                                                     \right),
  \label{ab}
\end{eqnarray}
where $s_{ij}$ and $ c_{ij}$ are sine and cosine of the lepton mixing angle $\theta_{ij}$,
   respectively.
The $U_{e i}  (i=1,2,3)$ satisfy
   $ \vert U_{e 1} \vert^2 + \vert U_{e 2} \vert^2 + \vert U_{e 3} \vert^2 = 1 $,
   and we shall restrict the Majorana phases, $ - \pi/2 < \beta, \alpha \le \pi/2 $,
   which are enough to investigate the variation of $ \vert m_{ee} \vert $.
   
The effective mass $ \vert m_{ee} \vert $ depends on seven parameters;
   two mixing angles $\theta_{12}$ and $ \theta_{13}$, three neutrino masses
   $m_1, m_2, m_3$, and two Majorana phases $\beta, \alpha$.
With regard to these seven parameters, the following four quantities are measured by 
   experiments; i.e., two mixing angles $\theta_{12}$, $ \theta_{13}$, and two mass
   squared differences $   \bigtriangleup m_{\odot}^2   \equiv  m_2^2 - m_1^2 $,
   $ \bigtriangleup  m_{\rm A}^2  \equiv  \vert m_3^2 - m_1^2 \vert
                \approx \vert m_3^2 - m_2^2 \vert $.
Two Majorana phases $\beta, \alpha$ (if Majorana particles) and the absolute neutrino
   mass scale (we can take the lightest neutrino mass $m_0$ as the absolute neutrino
   mass scale) are not measured, though an upper limit for the sum of the three light
   neutrino mass has been reported  \cite{rf:Ade,rf:HabTak}.
There remains two possibilities with respect to the order of $m_1, m_2$, and $m_3$,
   that is the normal mass ordering ($ m_3 >m_2>m_1 $) and inverted mass ordering
   ($ m_2 >m_1>m_3 $).
In the normal mass ordering, we take the lightest neutrino mass $ m_0 = m_1 $ as
   the absolute neutrino mass scale,
\begin{eqnarray}
   m_2 &=& \sqrt{ m_1^2 + \bigtriangleup m_{\odot}^2 },                  \nonumber  \\
   m_3 &=&  \sqrt{ m_1^2 +   \bigtriangleup  m_{\rm A}^2 }.
  \label{ac}
\end{eqnarray}
In the inverted mass ordering, we take the lightest neutrino mass $ m_0 = m_3 $ as
   the absolute neutrino mass scale,
\begin{eqnarray}
   m_2 &=& \sqrt{ m_3^2 + \bigtriangleup m_{\odot}^2 +  \bigtriangleup  m_{\rm A}^2 },                  \nonumber  \\
   m_1 &=&  \sqrt{ m_3^2 +   \bigtriangleup  m_{\rm A}^2 }.
  \label{ad}
\end{eqnarray}
The signal of $0\, \nu \beta \beta$ has not been detected, but an upper limit of the
   effective mass $ \vert m_{ee} \vert $ is obtained.
While information on Majorana phases $\beta, \alpha$ is obtained by $0\, \nu \beta \beta$
   experiments \cite{rf:MatTakFuk,rf:NunTevFun,rf:MinNunQui,rf:DodLyk}, it is notable
   that the other experiments measuring the absolute neutrino mass scale are also
   important to determine the phases $\beta, \alpha$ as discussed in
   Ref  \cite{rf:MinNunQui,rf:DodLyk}.
One of the most often used method for studying $0\, \nu \beta \beta$ is that,
   regarding the lightest neutrino mass $m_0$ as a free parameter, the predicted values
   of the effective mass $ \vert m_{ee} \vert $ are investigated for each given value of
   the lightest neutrino mass.
For instance, the authors in Ref \cite{rf:PasPet} calculated 1 $\sigma $ error on the
   predicted value of $ \vert m_{ee} \vert $ due to the uncertainties in the values of
   the neutrino oscillation parameters $\theta_{12}, \theta_{13}$,
   $ \bigtriangleup m_{\odot}^2 $, and $ \bigtriangleup  m_{\rm A}^2 $.
   
We here suppose that the lightest neutrino mass $m_0$ is measured,
\begin{equation}
  m_0 \pm \sigma ( m_0 ),
  \label{ae}
\end{equation}
where $\sigma ( m_0 )$ is 1 $ \sigma $ error on $m_0$.
If the normal mass ordering is assumed, we have $ m_0 = m_1 $;
   while if the inverted mass ordering, $ m_0 = m_3 $.
In this paper, we study contributions of the error on the lightest neutrino mass $m_0$
   to the uncertainty of the predicted value of the Majorana effective mass
   $ \vert m_{ee} \vert $ in the normal mass ordering case and in the inverted mass
   ordering case, respectively.
In order to predict the value of $ \vert m_{ee} \vert $ which depends on the Majorana
   phases $\beta$ and $ \alpha$, we need the values of the absolute neutrino mass
   scale (we can take the lightest neutrino mass $m_0$) and four oscillation parameters;
   $\theta_{12}, \theta_{13}$, $ \bigtriangleup m_{\odot}^2 $, and
   $ \bigtriangleup  m_{\rm A}^2 $.
These four oscillation parameters have been measured with relatively small errors,
   respectively.
Since we focus on the uncertainty $\sigma ( m_0 )$ of the lightest neutrino mass,
   we do not take into account the uncertainties of $\theta_{12}, \theta_{13}$,
   $ \bigtriangleup m_{\odot}^2 $, and $ \bigtriangleup  m_{\rm A}^2 $ in this article.
By use of the propagation of errors, the 1 $\sigma$ error on $ \vert m_{ee} \vert $
   due to the uncertainty of $m_0$ is obtained,
\begin{equation}
  \sigma (  \vert m_{ee} \vert )^2 
    = \left\{ \frac{ \partial \,  \vert m_{ee} \vert }{ \partial m_0 } \,  \sigma ( m_0 ) 
        \right\}^2.
  \label{af}
\end{equation}
The predicted value of $ \vert m_{ee} \vert $ depends on the Majorana phases
   $ \beta, \alpha $, and if the uncertainty $ \sigma ( \vert m_{ee} \vert ) $ is large,
   the dependence of $ \vert m_{ee} \vert $ on $ \beta$ and $\alpha $ is obscured.
To obtain some information on the Majorana phases $ \beta$ and $\alpha $ through
   the effective mass, the uncertainty $ \sigma ( \vert m_{ee} \vert ) $ should not be large.
We study how the uncertainty $ \sigma ( \vert m_{ee} \vert ) $ due to the uncertainty
   $\sigma ( m_0 )$ behaves in the $ \beta - \alpha $ plane, and how it changes 
   according to the size of the lightest neutrino mass $m_0$.
We also discuss the necessary condition for the relative error $ \sigma ( m_0 ) / m_0 $
   of the lightest neutrino mass to raise the possibility of probing
   the Majorana phases.
   
The paper is organized as follows.
In section 2, the normal mass ordering is assumed, so that the lightest neutrino mass
   $ m_0 =m_1 $.
The 1 $\sigma$ error $ \sigma ( \vert m_{ee} \vert ) $ on the predicted value of the
   effective mass $ \vert m_{ee} \vert $ is obtained by use of the propagation of
   errors, Eq.(\ref{af}).
The result of straightforward calculation of $ \sigma ( \vert m_{ee} \vert ) $ is so
   complicated that we introduce complex-valued quantities $P$ and $Q$
   in subsection 2.1.
Expressing $ \sigma ( \vert m_{ee} \vert ) $ by these complex-valued quantities
   $P$ and $Q$, one can analyze it easily.
We then investigate how $ \sigma ( \vert m_{ee} \vert ) $ behaves in the
   $ \beta - \alpha $ plane for each size of the lightest neutrino mass $m_1$.
In section 3, the inverted mass ordering is assumed, so that the lightest neutrino mass
   $ m_0 =m_3 $.
The 1 $\sigma$ error $ \sigma ( \vert m_{ee} \vert ) $ in the inverted mass ordering
   case is studied in the same manner as section 2.
In section 4, we discuss the necessary condition that the relative error of the lightest
   neutrino mass should satisfy in order to raise the possibility of probing the Majorana
   phases $ \beta, \alpha $.
Section 5 is devoted to conclusions.
%
%
%
%
%
\section{Normal mass ordering case  ( $ m_3 >m_2>m_1 $ )}
In this chapter, the normal mass ordering ( $ m_3 >m_2>m_1 $ ) is assumed and
   the lightest neutrino mass $ m_0 = m_1 $.
We assume that the lightest neutrino mas $m_1$ is measured,
\begin{equation}
  m_1 \pm \sigma ( m_1 ),
  \label{baa}
\end{equation}
where $  \sigma ( m_1 ) $ is the 1 $\sigma $ error on $m_1$.
The 1 $\sigma $ error $ \sigma ( \vert m_{ee} \vert ) $ of the predicted value of
   $ \vert m_{ee} \vert $ due to the uncertainty $  \sigma ( m_1 ) $ will be calculated
   and studied below.
As discussed in section 1, we do not consider the uncertainties of four oscillation
   parameters, $\theta_{12}, \theta_{13}$, $ \bigtriangleup m_{\odot}^2 $, and
   $ \bigtriangleup  m_{\rm A}^2 $ and therefore the $ \sigma ( \vert m_{ee} \vert ) $
   is obtained by Eq.(\ref{af}),
\begin{equation}
  \sigma ( \vert m_{ee} \vert )
    = \left\vert  \frac{ \partial \,  \vert m_{ee} \vert }{ \partial m_1 }  \right\vert
      \,  \sigma ( m_1 ).
  \label{bab}
\end{equation}
\subsection{ Calculation of $  \sigma ( \vert m_{ee} \vert ) $}
The explicit calculation of the factor $ \partial \,  \vert m_{ee} \vert / \partial m_1  $
   in the right-handed side of Eq.(\ref{bab}) shows
\newpage
\begin{eqnarray}
   \frac{ \partial \,  \vert m_{ee} \vert }{ \partial m_1 }
     &=&  \frac{ m_1 }{  \vert m_{ee} \vert }
            \left[  \vert U_{e 1} \vert^4 + \vert U_{e 2} \vert^4 + \vert U_{e 3} \vert^4
            + \left( { m_2 \over m_1 } +  { m_1 \over m_2 } \right)
                \vert U_{e 1} \vert^2  \vert U_{e 2} \vert^2 \cos 2 \alpha  \right.    \nonumber \\
    & & \hskip1.5cm 
           + \left( { m_3 \over m_1 } +  { m_1 \over m_3 } \right)
                \vert U_{e 3} \vert^2  \vert U_{e 1} \vert^2 \cos 2 \beta     \nonumber \\
    & & \hskip1.5cm  \left.
           + \left( { m_3 \over m_2 } +  { m_2 \over m_3 } \right)
                \vert U_{e 2} \vert^2  \vert U_{e 3} \vert^2 
                \left(  \cos 2 \alpha \cos 2 \beta + \sin 2 \alpha \sin 2 \beta \right)    \right],
   \label{bba}
\end{eqnarray}
where
\begin{eqnarray}
    \vert m_{ee} \vert 
   &=& \left[  ( m_1  \vert U_{e 1} \vert^2 )^2 +  ( m_2  \vert U_{e 2} \vert^2 )^2
        +  ( m_3  \vert U_{e 3} \vert^2 )^2      \right.       \nonumber \\
   & &  + 2  ( m_1  \vert U_{e 1} \vert^2 )( m_2 \vert U_{e 2} \vert^2 )\cos 2 \alpha
                                                                                                                      \nonumber \\
   & &  +  2  ( m_2  \vert U_{e 2} \vert^2 ) ( m_3  \vert U_{e 3} \vert^2 )
          \left(  \cos 2 \alpha \cos 2 \beta + \sin 2 \alpha \sin 2 \beta \right)          \nonumber \\
   & &  \left.   +  2  ( m_3  \vert U_{e 3} \vert^2 ) ( m_1  \vert U_{e 1} \vert^2 )
                             \cos 2 \beta \right]^{1 \over 2}.
  \label{bbb}
\end{eqnarray}
We prefer to study the behavior of $  \sigma ( \vert m_{ee} \vert ) $ in the
   $ \beta-\alpha $ plane analytically.
When it is difficult to study it analytically, the numerical calculations will be carried out
   in which the following reference values of four oscillation parameters are
   used \cite{rf:Gon},
\begin{eqnarray}
\bigtriangleup m_{\odot}^2 & = &  7.50 \times
                                   10^{-5} \, {\rm eV}^2,                  \nonumber  \\
    \bigtriangleup  m_{\rm A}^2  & = & 2.473 \times
                                   10^{-3} \, {\rm eV}^2,                  \nonumber  \\
   \sin^2 \theta_{12}  & = & 0.302,            \nonumber  \\
   \sin^2 \theta_{13}  & = & 0.0227.
      \label{bbc}
\end{eqnarray}
These reference values lead to
\begin{eqnarray}
    \vert U_{e 1} \vert^2 & = &  c_{1 2}^2 c_{1 3}^2 = 0.682,             \nonumber  \\
    \vert U_{e 2} \vert^2 & = &  s_{1 2}^2 c_{1 3}^2 = 0.295,             \nonumber  \\
    \vert U_{e 3} \vert^2 & = &  s_{1 3}^2 = 0.0227,
  \label{bbd}
\end{eqnarray}
and one has $  \vert U_{e 1} \vert^2 >  \vert U_{e 2} \vert^2 >>  \vert U_{e 3} \vert^2 $
   and $ \bigtriangleup  m_{\rm A}^2 >> \bigtriangleup m_{\odot}^2 $.
For an arbitrary value of $m_1$, the relation
   $ m_3  \vert U_{e 3} \vert^2 < m_2  \vert U_{e 2} \vert^2 $ is supposed in this section.

Although the analysis of the calculated result Eq.(\ref{bba}) of the factor
   $ \partial  \vert m_{ee} \vert / \partial m_1 $ is very hard, we can analyze it more
   easily by introducing the following two complex quantities $P$ and $Q$,
\begin{equation}
  P \equiv m_1  \vert U_{e 1} \vert^2 +m_2  \vert U_{e 2} \vert^2  e^{2 i \alpha}
                + m_3   \vert U_{e 3} \vert^2  e^{2 i \beta},
  \label{bbe}
\end{equation}
\begin{equation}
  Q \equiv {1 \over m_1 }  \vert U_{e 1} \vert^2 
        +{1 \over m_2}  \vert U_{e 2} \vert^2  e^{2 i \alpha}
        + {1 \over m_3}   \vert U_{e 3} \vert^2  e^{2 i \beta},
  \label{bbf}
\end{equation}
where $m_2$ and $m_3$ are given by Eq.(\ref{ac}).
Using the relation $ \partial P / \partial m_1 = m_1 Q $, one has
\begin{equation}
   \frac{ \partial \vert m_{ee} \vert }{ \partial m_1 }
    =  \frac{m_1}{ \vert P \vert } \, {\rm Re} [ P Q^* ]
    =  m_1  \vert Q^* \vert \, \frac{ {\rm Re} [ P Q^* ] }{ \vert P Q^*  \vert }
    =  m_1  \vert Q \vert \cos ( {\rm Arg} ( P Q^* ) ),
  \label{bbg}
\end{equation}
in which $  m_1  \vert Q \vert \ne 0 $ because of $ m_3 >m_2>m_1 $.
The absolute value of the factor $ \partial  \vert m_{ee} \vert / \partial m_1 $ satisfies
\begin{equation}
   \left\vert  \frac{ \partial \vert m_{ee} \vert }{ \partial m_1 }  \right\vert
     \le  \vert U_{e 1} \vert^2 + {m_1 \over m_2}  \vert U_{e 2} \vert^2 
    +  {m_1 \over m_3}  \vert U_{e 3} \vert^2
    < 1,
  \label{bbh}
\end{equation}
since $ \vert  \cos ( {\rm Arg} ( P Q^* ) ) \vert \le 1 $.
Now, we denote the maximum value of $  \sigma ( \vert m_{ee} \vert ) $
   in the $ \beta-\alpha $ plane as $  \sigma ( \vert m_{ee} \vert )_{ \rm max } $,
   and the minimum value as $  \sigma ( \vert m_{ee} \vert )_{ \rm min } $.
From Eq.(\ref{bbh}), the maximum value $  \sigma ( \vert m_{ee} \vert )_{ \rm max }  $
   in the $ \beta-\alpha $ plane is less than the value of $  \sigma ( m_1 ) $,
\begin{equation}
    \sigma ( \vert m_{ee} \vert )_{ \rm max }
     =   \left\{
           \vert U_{e 1} \vert^2 
          + {m_1 \over \sqrt{ m_1^2 + \bigtriangleup m_{\odot}^2 } }  \vert U_{e 2} \vert^2 
          +   {m_1 \over  \sqrt{ m_1^2 +   \bigtriangleup  m_{\rm A}^2 } }  
             \vert U_{e 3} \vert^2 \right\} \sigma( m_1 ).         \nonumber \\
     \label{bbi}
\end{equation}
The coefficient of $  \sigma ( m_1 ) $ in the right-handed side of Eq.(\ref{bbi}) is
   a monotone increasing function of the lightest neutrino mass $m_1$.
The dependence of $  \sigma ( \vert m_{ee} \vert )_{ \rm max } / \sigma (m_1) $
   on $m_1$ is shown in Fig.1 in which the lightest neutrino mass $ m_0 = m_1 $
   in the normal mass ordering.
%
%
%
\begin{figure}
     \centering
     \includegraphics[height=7cm]{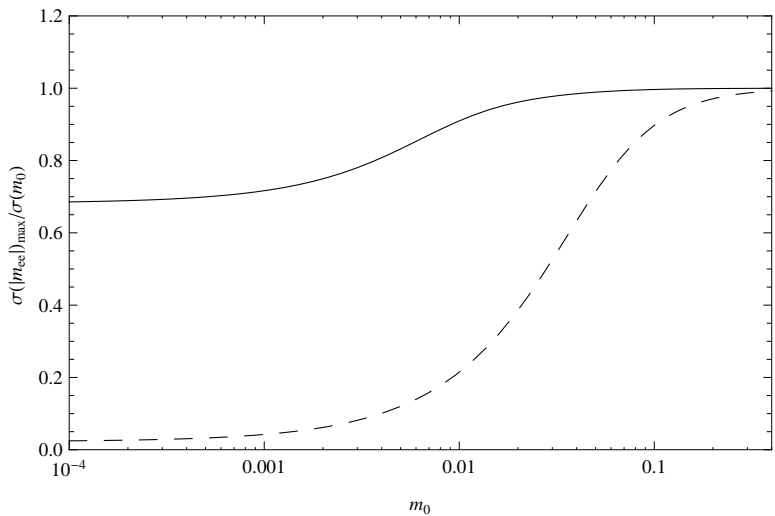}
    \caption{The value of $ \sigma ( \vert m_{ee} \vert )_{ \rm max } / \sigma (m_0) $
                  as a function of the lightest neutrino mass $m_0$  (in {\rm eV}).
                  The solid line represents the normal mass ordering case
                  ($m_0=m_1$), and the long-dashed line represents the inverted mass
                  ordering case ($m_0=m_3$), respectively.}
       \label{fig:1}
\end{figure}%
\noindent
As seen from the figure, the ratio $\sigma (\vert m_{ee} \vert )_{\rm max}/ \sigma(m_1)$
   is almost equal to 1 when the lightest neutrino mass $m_1$ is large, whereas it is
   equal to the value $   \vert U_{e 1} \vert^2 \approx 0.68 $ when $m_1$ is
   approximately zero.
The case of the inverted mass ordering in Fig.1 will be discussed in section 3.
The minimum value $ \sigma ( \vert m_{ee} \vert )_{ \rm min } $ in the
   $ \beta-\alpha $ plane can take the value zero if
   ${\rm Re} [ P Q^* ] / \vert P Q^*  \vert =  \cos ( {\rm Arg} ( P Q^* ) ) =0$
   is realized.
\subsection{ Behavior of $ \sigma( \vert m_{ee} \vert ) $ according
	                 to the size of $m_1$. }
In this subsection, the behavior of $\sigma( \vert m_{ee} \vert)$
   in the $ \beta - \alpha $ plane is investigated according to the size of the lightest
   neutrino mass $m_1$ in the normal mass ordering case.
The expression of the maximum value $ \sigma (\vert m_{ee} \vert )_{\rm max}$
   is obtained in the preceding subsection.
We are especially interested in the case where $\sigma( \vert m_{ee} \vert)$
   takes the value zero or approximately zero in the $ \beta - \alpha $ plane,
   since the uncertainty of the predicted value of
   $ \vert m_{ee} \vert $ due to the uncertainty of the lightest
   neutrino mass becomes zero in that case.
The behavior of $\sigma( \vert m_{ee} \vert)$ will be studied according to the size of
   $m_1$ in the following (A)〜(D).
\subsubsection{ (A): $ m_1 \rightarrow 0 $}
In this case, the upper limit of the measured value of the lightest neutrino mass is
   almost zero.
Although $ \sigma (m_1) \rightarrow 0 $ would be impossible in practice, we shall
   study this case because of the characteristic behavior of
   $\partial \vert m_{ee} \vert / \partial m_1$ in the limit $ m_1 \rightarrow 0 $,
\begin{equation}
 \frac{ \partial \vert m_{ee} \vert }{ \partial m_1 }
   \rightarrow  \vert U_{e 1} \vert^2 
      \frac{  {\rm Re} (P) }
       {
         \sqrt{  \{  {\rm Re} (P) \}^2 +  \{  {\rm Im} (P) \}^2 }  
                                                                                          }.
  \label{bca}
\end{equation}
On the plane curve in the $ \beta - \alpha $ plane,
\begin{equation}
   {\rm Re} ( P ) =  \sqrt{ \bigtriangleup m_{\odot}^2 } \,  \vert U_{e 2} \vert^2
          \cos 2 \alpha
         +   \sqrt{ \bigtriangleup  m_{\rm A}^2 } \,
               \vert U_{e 3} \vert^2 \cos 2 \beta = 0,
  \label{bcb}
\end{equation}
or
\begin{equation}
   \cos 2 \alpha + 0.442 \cos 2 \beta =0,
  \label{bcc}
\end{equation}
the factor ${ \partial \vert m_{ee} \vert }/{ \partial m_1 } $ in Eq.(\ref{bca}) takes the
   minimum value zero,
\begin{equation}
   \left(  \frac{ \partial \vert m_{ee} \vert }{ \partial m_1 } \right)_{\rm min }
   =  \left(  \frac{ \partial \vert m_{ee} \vert }{ \partial m_1 } \right)_{{\rm Re} ( P )=0  } 
    =0.
  \label{bcd}
\end{equation}
On the other side, on the plane curve in the $ \beta - \alpha $ plane,
\begin{equation}
   {\rm Im} ( P ) =  \sqrt{ \bigtriangleup m_{\odot}^2 } \,  \vert U_{e 2} \vert^2
          \sin 2 \alpha
         +   \sqrt{ \bigtriangleup  m_{\rm A}^2 } \,
               \vert U_{e 3} \vert^2 \sin 2 \beta = 0,
  \label{bce}
\end{equation}
or
\begin{equation}
   \sin 2 \alpha + 0.442 \sin 2 \beta =0,
  \label{bcf}
\end{equation}
the factor takes the maximum value,
\begin{equation}
   \left(  \frac{ \partial \vert m_{ee} \vert }{ \partial m_1 } \right)_{\rm max }
   =  \left(  \frac{ \partial \vert m_{ee} \vert }{ \partial m_1 } \right)_{{\rm Im} ( P )=0  } 
   =  \vert U_{e 1} \vert^2 = 0.682.
  \label{bcg}
\end{equation}
\subsubsection{ (B): $  \vert U_{e 1} \vert^2 / m_1  \gg  \vert U_{e 2} \vert^2 / m_2 $
  \hskip0.1cm  ( $ m_1 \ll   \sqrt{ \bigtriangleup m_{\odot}^2 } \,  \vert U_{e 1} \vert^2 /
                           \vert U_{e 2} \vert^2  \sim  0.020 \, {\rm eV} )  $ }
For $ m_1 \ll  0.020 \, {\rm eV}, $ $Q$ can be approximated by
\begin{equation}
     Q \approx {1 \over m_1 }  \vert U_{e 1} \vert^2,
  \label{bch}
\end{equation}
which leads to
\begin{equation}
 \frac{ \partial \vert m_{ee} \vert }{ \partial m_1 }
   \approx ( m_1 Q ) \, \frac{   {\rm Re} (P) }{ \vert P \vert }
   \approx  \vert U_{e 1} \vert^2 
     \, \frac{  {\rm Re} (P) }
       {
         \sqrt{  \{  {\rm Re} (P) \}^2 +  \{  {\rm Im} (P) \}^2 }  
                                                                                          }.
  \label{bci}
\end{equation}
On the plane curve in the $ \beta - \alpha $ plane,
\begin{equation}
     {\rm Re} ( P ) =   m_1  \vert U_{e 1} \vert^2 
             +  m_2  \vert U_{e 2} \vert^2   \cos 2 \alpha
             +  m_3  \vert U_{e 3} \vert^2  \cos 2 \beta = 0,
  \label{bcj}
\end{equation}
or
\begin{equation}
    0.682 \, m_1  
    + 0.295 \, \sqrt{ m_1^2 + \bigtriangleup m_{\odot}^2 } \,   \cos 2 \alpha
         +  0.0227 \,  \sqrt{ m_1^2 + \bigtriangleup  m_{\rm A}^2 } \,
               \cos 2 \beta = 0,
  \label{bck}
\end{equation}
the factor becomes approximately zero
   ${ \partial \vert m_{ee} \vert }/{ \partial m_1 } \approx 0$, and we have
\begin{equation}
    \sigma ( \vert m_{ee} \vert )_{ \rm min }
    =   \sigma ( \vert m_{ee} \vert )_{ {\rm Re} ( P ) = 0  } 
    \approx 0.
  \label{bcl}
\end{equation}
From the equation of the above plane curve Eq.(\ref{bcj}), $\cos 2 \alpha$ can
   be expressed as \\
   $  \cos 2 \alpha
          = ( - m_1  \vert U_{e 1} \vert^2 - m_3  \vert U_{e 3} \vert^2  \cos 2 \beta )
                 / ( m_2  \vert U_{e 2} \vert^2 ) $,
   which leads to
\begin{equation}
    \frac{  - m_1  \vert U_{e 1} \vert^2 - m_3  \vert U_{e 3} \vert^2 }{ m_2  \vert U_{e 2} \vert^2 }
      \le  \cos 2 \alpha
      \le   \frac{  - m_1  \vert U_{e 1} \vert^2 + m_3  \vert U_{e 3} \vert^2 }
                    { m_2  \vert U_{e 2} \vert^2 }.
   \label{bcm}
\end{equation}
If $m_1$ satisfies
   $ ( - m_1  \vert U_{e 1} \vert^2 + m_3  \vert U_{e 3} \vert^2  )
            / ( m_2  \vert U_{e 2} \vert^2 )  < -1$,
   that is, $ 0.0063 {\rm eV} < m_1 $, two Majorana phases $ ( \beta, \alpha ) $
   satisfying $ {\rm Re} ( P ) = 0 $ do not exist.
Actually, the numerical calculation shows that, for 
   $ 0.0063 \,
    {\rm eV}  \mathrel{ \rlap{\raise 0.511ex \hbox{$<$}}{\lower 0.511ex \hbox{$\sim$}}} m_1 $,
   there exists no plane curve in the $ \beta - \alpha $ plane which satisfies
   $  \sigma ( \vert m_{ee} \vert )=0 $.

On the other hand, on the plane curve in the $ \beta - \alpha $ plane,
\begin{equation}
     {\rm Im} ( P ) =   m_2  \vert U_{e 2} \vert^2   \sin 2 \alpha
        +  m_3  \vert U_{e 3} \vert^2   \sin 2 \beta = 0,
  \label{bcn}
\end{equation}
or
\begin{equation}
  0.295 \, \sqrt{ m_1^2 + \bigtriangleup m_{\odot}^2 } \, \sin 2 \alpha
         +   0.0227 \, \sqrt{ m_1^2 + \bigtriangleup  m_{\rm A}^2 } \, \sin 2 \beta = 0,
  \label{bco}
\end{equation}
the factor ${ \partial \vert m_{ee} \vert }/{ \partial m_1 }$ takes the maximum value
   $  \vert U_{e 1} \vert^2 $ and thereby,
\begin{equation}
    \sigma ( \vert m_{ee} \vert )_{ \rm max }
    =   \sigma ( \vert m_{ee} \vert )_{ {\rm Im} ( P )=0  } 
    \approx  \vert U_{e 1} \vert^2 \sigma ( m_1 ) = 0.682 \,  \sigma ( m_1 ).
  \label{bcp}
\end{equation}
\subsubsection{ (C):  $ m_2  \vert U_{e 2} \vert^2  \gg   m_3  \vert U_{e 3} \vert^2 $
    and  $   m_1^2
                \mathrel{ \rlap{\raise 0.511ex \hbox{$<$}}{\lower 0.511ex \hbox{$\sim$}}}
                  \bigtriangleup  m_{\rm A}^2  $
  \hskip0.1cm ( $  0.01 \, {\rm eV} 
                        \mathrel{ \rlap{\raise 0.511ex \hbox{$<$}}{\lower 0.511ex \hbox{$\sim$}}}
                        m_1 
                       \mathrel{ \rlap{\raise 0.511ex \hbox{$<$}}{\lower 0.511ex \hbox{$\sim$}}}
                        0.05 \, {\rm eV}   $ )   }
In this case, we obtain the following relation,
   $  \vert U_{e 2} \vert^2  \gg  (m_3 / m_2)  \vert U_{e 3} \vert^2 
         >   (m_2 / m_3) \vert U_{e 3} \vert^2 $,
   which allows us to use the approximation,
\begin{eqnarray}
  P  & \approx &  m_1  \vert U_{e 1} \vert^2 +m_2  \vert U_{e 2} \vert^2  e^{2 i \alpha},
                                                           \nonumber \\
  Q  & \approx &  {1 \over m_1 }  \vert U_{e 1} \vert^2 
        +{1 \over m_2}  \vert U_{e 2} \vert^2  e^{2 i \alpha}.
  \label{bcq}
\end{eqnarray}
The factor ${ \partial \vert m_{ee} \vert }/{ \partial m_1 }$ is then approximated,
\begin{equation}
   \frac{ \partial \vert m_{ee} \vert }{ \partial m_1 }  \,
    \approx  \,
        \frac{m_1  \left\{  \vert U_{e 1} \vert^4 +  \vert U_{e 2} \vert^4 
                  + \left( {m_2 \over m_1} +  {m_1 \over m_2} \right)
                      \vert U_{e 1} \vert^2  \vert U_{e 2} \vert^2 \cos 2 \alpha \right\}  }
               { ( m_1  \vert U_{e 1} \vert^2 )^2 +  ( m_2  \vert U_{e 2} \vert^2 )^2
                   + 2  ( m_1  \vert U_{e 1} \vert^2 ) ( m_2  \vert U_{e 2} \vert^2 ) \cos 2 \alpha }.
     \label{bcr}
\end{equation}
The factor ${ \partial \vert m_{ee} \vert }/{ \partial m_1 }$ scarcely depends on the
   Majorana phase $\beta$, and is a monotone decreasing function of the Majorana
   phase $\alpha \, ( 0 \leq \alpha \leq \pi / 2 ) $.
The ${ \partial \vert m_{ee} \vert }/{ \partial m_1 }$ takes the maximum value
   $\vert U_{e 1} \vert^2 + (m_1 / m_2)  \vert U_{e 2} \vert^2 + O( \vert U_{e 3} \vert^2 )$
   at $ \alpha = 0$, and the minimum value
   $\vert U_{e 1} \vert^2 - (m_1 / m_2)  \vert U_{e 2} \vert^2 + O( \vert U_{e 3} \vert^2 )$
   at $ \alpha = \pi/2$.
Namely, $\sigma ( \vert m_{ee} \vert ) $ scarcely depends on $\beta$, and is
   a monotone decreasing function of $\alpha \, ( 0 \leq \alpha \leq \pi / 2 ) $.
The $\sigma ( \vert m_{ee} \vert ) $ takes the maximum value at $ \alpha = 0$,
\begin{eqnarray}
    \sigma ( \vert m_{ee} \vert )_{ \rm max }
    & = &   \sigma ( \vert m_{ee} \vert )_{ \alpha =0 }            \nonumber \\
    &  \approx &   \left\{ \vert U_{e 1} \vert^2 + {m_1 \over m_2}  \vert U_{e 2} \vert^2 \right \}
                     \sigma ( m_1 )                                                    \nonumber \\
    & = &  \left\{ 0.682 + 0.295 \,  \frac{m_1}{ \sqrt{ m_1^2 + \bigtriangleup m_{\odot}^2 }  }
                  \right \}  \sigma ( m_1 ),
  \label{bcs}
\end{eqnarray}
and the minimum value at $\alpha=\pi / 2$,
\begin{eqnarray}
    \sigma ( \vert m_{ee} \vert )_{ \rm min }
    & = &   \sigma ( \vert m_{ee} \vert )_{ \alpha = \pi /2 }            \nonumber \\
    &  \approx &   \left\{ \vert U_{e 1} \vert^2 - {m_1 \over m_2}  \vert U_{e 2} \vert^2 \right \}
                     \sigma ( m_1 )                                                    \nonumber \\
    & = &  \left\{ 0.682 - 0.295 \,  \frac{m_1}{ \sqrt{ m_1^2 + \bigtriangleup m_{\odot}^2 }  }
                  \right \}  \sigma ( m_1 ),
  \label{bct}
\end{eqnarray}
where we have neglected the term of the order $ O( \vert U_{e 3} \vert^2 ) $
   in ${ \partial \vert m_{ee} \vert }/{ \partial m_1 }$.
\subsubsection{ (D): $ m_1^2 \gg { \bigtriangleup  m_{\rm A}^2 } $
                          such that $ m_1 \approx m_2 \approx m_3 $
                          \hskip0.1cm  $( m_1^2  \gg ( 0.05 {\rm eV} )^2 )$   }
In this case, using the relation $ Q \approx  P / m_1^2 $, one has
\begin{equation}
     \frac{ \partial \, \vert m_{ee} \vert }{ \partial \, m_1 } 
     \approx \frac{  \vert m_{ee} \vert }{m_1} 
     \approx  \vert \vert U_{e 1} \vert^2 + \vert U_{e 2} \vert^2  e^{2 i \alpha}
                + \vert U_{e 3} \vert^2  e^{2 i \beta} \vert,
  \label{bcu}
\end{equation}
or
\begin{equation}
   \frac{ \sigma ( \vert m_{ee} \vert ) }{ \vert m_{ee} \vert }
   \approx  \frac{ \sigma ( m_1 ) }{ m_1 },
  \label{bcv}
\end{equation}
which means that the relative error of the effective mass $  \vert m_{ee} \vert $ has
   the same magnitude as that of the lightest neutrino mass $m_1$.
If the relative error of $m_1$ is fixed, $ \sigma ( \vert m_{ee} \vert ) $ is proportional
   to $  \vert m_{ee} \vert $.
The $  \vert m_{ee} \vert $ takes the maximum value
   $ m_1 \{  \vert U_{e 1} \vert^2 + \vert U_{e 2} \vert^2 + \vert U_{e 3} \vert^2 \}=m_1 $
   in the $ \beta - \alpha $ plane at the point $ (\beta, \alpha) = (0, 0) $,
   therefore $ \sigma ( \vert m_{ee} \vert ) $ takes the maximum value,
\begin{equation}
    \sigma ( \vert m_{ee} \vert )_{ \rm max }
    =   \sigma ( \vert m_{ee} \vert )_{ (\beta, \alpha)=(0,0) } 
       \approx  \sigma ( m_1 ),
  \label{bcw}
\end{equation}
at the same point $ (\beta, \alpha) = (0, 0) $.
On the other side, $  \vert m_{ee} \vert $ takes the minimum value
   $ m_1 \{  \vert U_{e 1} \vert^2 - \vert U_{e 2} \vert^2 - \vert U_{e 3} \vert^2 \} $
   in the $ \beta - \alpha $ plane at the point $ (\beta, \alpha) = (\pi/2, \pi/2) $,
   therefore $ \sigma ( \vert m_{ee} \vert ) $ takes the minimum value,
\begin{eqnarray}
    \sigma ( \vert m_{ee} \vert )_{ \rm min }
    & = &   \sigma ( \vert m_{ee} \vert )_{ (\beta, \alpha)=( \pi/2, \pi/2) }          \nonumber \\
    &  \approx &   \left\{ \vert U_{e 1} \vert^2 - \vert U_{e 2} \vert^2  - \vert U_{e 3} \vert^2
                                        \right \}   \sigma ( m_1 )                                       \nonumber \\
    & = & 0.364 \,  \sigma ( m_1 ),
  \label{bcx}
\end{eqnarray}
at the same point $ (\beta, \alpha) = (\pi/2, \pi/2) $.
\section{Inverted mass ordering case; ( $ m_2 >m_1>m_3 $ ) }
In this chapter, the inverted mass ordering ( $ m_2 >m_1>m_3 $ ) is supposed and
   the lightest neutrino mass $ m_0 = m_3 $.
We assume that the lightest neutrino mass $m_3$ is measured,
\begin{equation}
  m_3 \pm \sigma ( m_3 ),
  \label{caa}
\end{equation}
where $  \sigma ( m_3 ) $ is the 1 $\sigma $ error on $m_3$.
The 1 $\sigma $ error $ \sigma ( \vert m_{ee} \vert ) $ of the predicted value of
   $ \vert m_{ee} \vert $ due to the uncertainty $  \sigma ( m_3 ) $ is obtained
   from Eq.(\ref{af}),
\begin{equation}
  \sigma ( \vert m_{ee} \vert )
    = \left\vert  \frac{ \partial \,  \vert m_{ee} \vert }{ \partial m_3 }  \right\vert
      \,  \sigma ( m_3 ),
  \label{cab}
\end{equation}
where we have not considered the uncertainties of four oscillation
   parameters, $\theta_{12}, \theta_{13}$, $ \bigtriangleup m_{\odot}^2 $, and
   $ \bigtriangleup  m_{\rm A}^2 $ as in section 2.
\subsection{Calculation of $  \sigma ( \vert m_{ee} \vert ) $}
The explicit calculation of the factor $  \partial \,  \vert m_{ee} \vert / \partial m_3  $
   shows that the expression of it is very complicated as in the normal mass ordering
   case, and we shall not present here the calculated result.
As in section 2, we prefer to study the behavior of $\sigma ( \vert m_{ee} \vert )$
   in the $ \beta-\alpha $ plane analytically also in the inverted mass ordering case.
When it is difficult to study it analytically, the numerical calculations will be carried out
   in which the reference values Eq.(\ref{bbc}) and Eq.(\ref{bbd}) are also used
   in this chapter.
In the inverted mass ordering, the relation 
   $  m_1 \vert U_{e 1} \vert^2 >  m_2 \vert U_{e 2} \vert^2  \gg  m_3\vert U_{e 3} \vert^2$
   holds for an arbitrary value of the lightest neutrino mass $m_3$.
Furthermore, the following relation is satisfied,
\begin{equation}
  1 < { m_2 \over m_1 }
     = \sqrt{ 1 + \frac{  \bigtriangleup  m_{\odot}^2  }{  m_3^2 +   \bigtriangleup  m_{\rm A}^2 } }
     <   \sqrt{ 1 + \frac{  \bigtriangleup  m_{\odot}^2  }{  \bigtriangleup  m_{\rm A}^2 } }
     \approx 1.015,
  \label{cba}
\end{equation}
thereby, we use the approximation $  m_1 \approx m_2 $.
While the expression of $ \partial \,  \vert m_{ee} \vert / \partial m_3$ calculated
   explicitly is very complicated, we can analyze $\partial \,  \vert m_{ee} \vert / \partial m_3$
   more easily by introducing the following complex quantities $P$ and $Q$,
   as in section 2,
\begin{equation}
  P \equiv m_1  \vert U_{e 1} \vert^2 +m_2  \vert U_{e 2} \vert^2  e^{2 i \alpha}
                + m_3   \vert U_{e 3} \vert^2  e^{2 i \beta},
  \label{cbb}
\end{equation}
\begin{equation}
  Q \equiv {1 \over m_1 }  \vert U_{e 1} \vert^2 
        +{1 \over m_2}  \vert U_{e 2} \vert^2  e^{2 i \alpha}
        + {1 \over m_3}   \vert U_{e 3} \vert^2  e^{2 i \beta},
  \label{cbc}
\end{equation}
where $m_1$ and $m_2$ are given by Eq.(\ref{ad}).
It should be noted that $( m_1, m_2, m_3)$ of the inverted mass ordering case
   is different from $( m_1, m_2, m_3)$ of the normal mass ordering case.
Using the relation, $ \partial P / \partial m_3 = m_3 Q $, we have \\
\begin{equation}
   \frac{ \partial \vert m_{ee} \vert }{ \partial m_3 }
    =  \frac{m_3}{ \vert P \vert } \, {\rm Re} [ P Q^* ]
    =  m_3  \vert Q \vert \cos ( {\rm Arg} ( P Q^* ) ),
  \label{cbd}
\end{equation}
which leads to
\begin{equation}
   \left\vert   \frac{ \partial \vert m_{ee} \vert }{ \partial m_3 }   \right\vert
     \le   {m_3 \over m_1}  \vert U_{e 1} \vert^2 + {m_3 \over m_2}  \vert U_{e 2} \vert^2 
    +  \vert U_{e 3} \vert^2
    < 1.
  \label{cbe}
\end{equation}

Here, we denote the maximum value of $  \sigma ( \vert m_{ee} \vert ) $ in the
   $ \beta-\alpha $ plane as  $  \sigma ( \vert m_{ee} \vert )_{ \rm max } $,
   and the minimum value as $  \sigma ( \vert m_{ee} \vert )_{ \rm min } $.
From Eq.(\ref{cbe}), $  \sigma ( \vert m_{ee} \vert )_{ \rm max }  $ is less
   than $  \sigma ( m_3 ) $,
\begin{equation}
    \sigma ( \vert m_{ee} \vert )_{ \rm max }
      =  \left\{  {m_3 \over  \sqrt{ m_3^2 +   \bigtriangleup  m_{\rm A}^2 } } 
           \vert U_{e 1} \vert^2 
      + {m_3 \over \sqrt{ m_3^2 + \bigtriangleup m_{\odot}^2 +  \bigtriangleup  m_{\rm A}^2 } }
          \vert U_{e 2} \vert^2 
          +  \vert U_{e 3} \vert^2 \right\} \sigma( m_3 ).
     \label{cbf}
\end{equation}
The coefficient of $  \sigma ( m_3) $ in the right-handed side of Eq.(\ref{cbf}) is
   a monotone increasing function of the lightest neutrino mass $m_3$.
The dependence of $  \sigma ( \vert m_{ee} \vert )_{ \rm max } / \sigma (m_3) $
   on $m_3$ is shown in Fig.1 in which the lightest neutrino mass $ m_0 = m_3 $
   in the inverted mass ordering.
In the inverted mass ordering case, the ratio of the maximum value of
   $  \sigma ( \vert m_{ee} \vert ) $ in the $ \beta-\alpha $ plane to $  \sigma ( m_3) $,
   $\sigma (\vert m_{ee} \vert )_{\rm max}/ \sigma(m_3)$,
   is almost equal to 1 when the lightest neutrino mass $m_3$ is large
   such as $m_3^2 \gg  \bigtriangleup  m_{\rm A}^2$.
When the value of $m_3$ decreases, the value of the ratio
   $\sigma (\vert m_{ee} \vert )_{\rm max}/ \sigma(m_3)$
   decreases rapidly and it becomes $\vert U_{e 3} \vert^2 \approx 0.02$
   if $m_3$ is approximately zero.
On the other hand, in the normal mass ordering case discussed in section 2,
   the ratio $\sigma (\vert m_{ee} \vert )_{\rm max}/ \sigma(m_1)$
   is almost 1 when the lightest neutrino mass $m_1$ is large such as
   $ m_1^2 \gg  \bigtriangleup  m_{\rm A}^2 $.
When the value of $m_1$ decreases, however, the value of
   $\sigma (\vert m_{ee} \vert )_{\rm max}/ \sigma(m_1)$ decreases gradually and
   it becomes $\vert U_{e 1} \vert^2 \approx 0.68 $ if  $m_1$ is approximately zero.
In the case of large value of the lightest neutrino mass
   $m_0^2 \gg \bigtriangleup  m_{\rm A}^2 $, the above results can be easily understood
   since the difference between the normal mass ordering and the inverted mass ordering
   disappears when the neutrino masses $m_1$, $m_2$, and $m_3$ are degenerate.
If the lightest neutrino mass $m_0$ is small, one can see from Fig.1 that the value of
   $ \sigma ( \vert m_{ee} \vert )_{ \rm max } / \sigma (m_0) $ in the inverted mass
   ordering case is much less than that in the normal mass ordering case.

The minimum value $ \sigma ( \vert m_{ee} \vert )_{ \rm min } $ in the
   $ \beta-\alpha $ plane can take the value zero if either
   ${\rm Re} [ P Q^* ] / \vert P Q^*  \vert =  \cos ( {\rm Arg} ( P Q^* ) ) =0$
   or $  \vert Q  \vert = 0  $ is realized.
There is a possibility that $  \vert Q  \vert $ becomes zero in the inverted mass
   ordering case, while that is impossible in the normal mass ordering case.
The possibility $  \vert Q  \vert = 0  $ will be also considered below.
\subsection{ Behavior of $ \sigma( \vert m_{ee} \vert ) $ according
	                 to the size of $m_3$. }
In this subsection, the behavior of $ \sigma( \vert m_{ee} \vert ) $ in the $ \beta - \alpha $
   plane is investigated according to the size of the lightest neutrino mass $m_3$
   in the inverted mass ordering case.
The expression of the maximum value $\sigma ( \vert m_{ee} \vert )_{ \rm max }$
   is obtained in the preceding subsection.
As discussed in subsection 2.2, we are especially interested in the case where
   $ \sigma( \vert m_{ee} \vert ) $ takes the value zero or approximately zero in the
   $ \beta - \alpha $ plane.
In the inverted mass ordering case, $ \sigma( \vert m_{ee} \vert ) $ can be zero
   not only if $ \cos ( {\rm Arg} ( P Q^* ) )=0 $ but also if $  \vert Q  \vert = 0  $.
The behavior of $ \sigma( \vert m_{ee} \vert ) $ will be studied according to the size
   of $m_3$ in the following (A)〜(F).
\subsubsection{ (A): $ m_3 \rightarrow 0 $}
In the limit $ m_3 \rightarrow 0 $, the factor $  \partial \vert m_{ee} \vert / \partial m_3  $
   becomes
\begin{equation}
 \frac{ \partial \vert m_{ee} \vert }{ \partial m_3 }
   =  m_3 \vert Q \vert \,
       \frac{ {\rm Re} [  P Q^*  ] }{ \vert  P Q^* \vert }  
   \rightarrow   \vert U_{e 3} \vert^2 
      \frac{  {\rm Re} ( m_3 Q^* P) }
       {
         \sqrt{  \{  {\rm Re} ( m_3 Q^* P) \}^2 +  \{  {\rm Im} ( m_3 Q^* P) \}^2 }  
                                                                                          }.
  \label{cca}
\end{equation}
On the plane curve in the $ \beta - \alpha $ plane,
\begin{equation}
    { 1 \over  \vert U_{e 3} \vert^2 }   {\rm Re} (  m_3 Q^* P )
    =  \sqrt{ \bigtriangleup m_{\rm A}^2 } \,  \vert U_{e 1} \vert^2
          \cos ( -2 \beta )
         +   \sqrt{ \bigtriangleup  m_{\rm A}^2 + \bigtriangleup m_{\odot}^2 } \,
               \vert U_{e 2} \vert^2 \cos( 2 \alpha - 2 \beta ) = 0,
  \label{ccb}
\end{equation}
or
\begin{equation}
   \cos (- 2 \beta ) + 0.44 \cos ( 2 \alpha -2 \beta ) =0,
  \label{ccc}
\end{equation}
the factor ${ \partial \vert m_{ee} \vert }/{ \partial m_3 } $ in Eq.(\ref{cca}) takes the
   minimum value zero,
\begin{equation}
   \left(  \frac{ \partial \vert m_{ee} \vert }{ \partial m_3 } \right)_{\rm min }
   =  \left(  \frac{ \partial \vert m_{ee} \vert }{ \partial m_3 }
                                   \right)_{{\rm Re} (  m_3 Q^* P )=0  } 
    =0.
  \label{ccd}
\end{equation}
On the other hand, on the plane curve in the $ \beta - \alpha $ plane,
\begin{equation}
   {\rm Im} (  m_3 Q^* P ) =  \sqrt{ \bigtriangleup m_{\rm A}^2 } \,  \vert U_{e 1} \vert^2
          \sin ( -2 \beta )
         +   \sqrt{ \bigtriangleup  m_{\rm A}^2 + \bigtriangleup m_{\odot}^2 } \,
               \vert U_{e 2} \vert^2 \sin ( 2 \alpha - 2 \beta ) = 0,
  \label{cce}
\end{equation}
or
\begin{equation}
   \sin (- 2 \beta ) + 0.44 \sin ( 2 \alpha -2 \beta ) =0,
  \label{ccf}
\end{equation}
the factor ${ \partial \vert m_{ee} \vert }/{ \partial m_3 } $ takes the maximum value,
\begin{equation}
   \left(  \frac{ \partial \vert m_{ee} \vert }{ \partial m_3} \right)_{\rm max }
   =  \left(  \frac{ \partial \vert m_{ee} \vert }{ \partial m_3 } \right)_{{\rm Im} ( m_3 Q^* P )=0  } 
   =  \vert U_{e 3} \vert^2 = 0.0227.
  \label{ccg}
\end{equation}
\subsubsection{ (B):  $  \vert U_{e 3} \vert^2 / m_3  \gg 
                                          \vert U_{e 1} \vert^2 / m_1 $
       $  \hskip0.1cm  ( m_3 \ll  \sqrt{ \bigtriangleup m_{\rm A}^2 } \, \vert U_{e 3} \vert^2 /
                           \vert U_{e 1} \vert^2  \sim  0.002 \, {\rm eV} )  $ }
For $ m_3 \ll   0.002 \, {\rm eV} $, $Q$ can be approximated by
\begin{equation}
     Q \approx {1 \over m_3 }  \vert U_{e 3} \vert^2 e^{2 i \beta},
  \label{cch}
\end{equation}
which leads to
\begin{equation}
 \frac{ \partial \vert m_{ee} \vert }{ \partial m_3 }
     \approx  \vert U_{e 3} \vert^2 
      \frac{  {\rm Re} ( m_3 Q^* P) }
       {
         \sqrt{  \{  {\rm Re} ( m_3 Q^* P) \}^2 +  \{  {\rm Im} ( m_3 Q^* P) \}^2 }  
                                                                                          }.
  \label{cci}
\end{equation}
On the plane curve,
\begin{eqnarray}
       & &  { 1 \over  \vert U_{e 3} \vert^2 }  {\rm Re} ( m_3 Q^* P )   \nonumber \\
       & \approx &   m_1 \vert U_{e 1} \vert^2   \cos ( - 2 \beta ) 
         +   m_2 \vert U_{e 2} \vert^2 \cos ( -2 \beta + 2 \alpha )
        +  m_3 \vert U_{e 3} \vert^2   = 0,
  \label{ccj}
\end{eqnarray}
or
\begin{equation}
    0.682 \, \sqrt{ \bigtriangleup  m_{\rm A}^2 } \, \cos ( - 2 \beta )
    + 0.295 \, \sqrt{ \bigtriangleup  m_{\rm A}^2 + \bigtriangleup m_{\odot}^2 } \, 
                         \cos ( -2 \beta + 2 \alpha )
    + 0.0227 \, m_3 = 0,
  \label{cck}
\end{equation}
the factor becomes approximately zero
   ${ \partial \vert m_{ee} \vert }/{ \partial m_3 } \approx 0$.
The inequality $ m_3^2 \ll  \bigtriangleup  m_{\rm A}^2 $ has been used in
   driving Eq.(\ref{cck}).
We therefore have
\begin{equation}
    \sigma ( \vert m_{ee} \vert )_{ \rm min }
    =   \sigma ( \vert m_{ee} \vert )_{ {\rm Re} ( m_3 Q^* P ) = 0  } 
    \approx 0.
  \label{ccl}
\end{equation}
On the other side, on the plane curve,
\begin{equation}
      { 1 \over  \vert U_{e 3} \vert^2 } {\rm Im} ( m_3 Q^* P )
       =   m_1 \vert U_{e 1} \vert^2  \sin (- 2 \beta )
        +  m_2 \vert U_{e 2} \vert^2  \sin ( { - 2 \beta + 2 \alpha } )= 0,
  \label{ccm}
\end{equation}
or
\begin{equation}
     0.682 \, \sqrt{ \bigtriangleup  m_{\rm A}^2 } \, \sin ( - 2 \beta )
    + 0.295 \, \sqrt{ \bigtriangleup  m_{\rm A}^2 + \bigtriangleup m_{\odot}^2 } \, 
                       \sin ( -2 \beta + 2 \alpha ) = 0,
  \label{ccn}
\end{equation}
the factor ${ \partial \vert m_{ee} \vert }/{ \partial m_3 }$ takes the maximum value
   $  \vert U_{e 3} \vert^2 $.
We have again used the inequality $ m_3^2 \ll  \bigtriangleup  m_{\rm A}^2 $
   in deriving Eq.(\ref{ccn}).
The maximum value of $  \sigma ( \vert m_{ee} \vert ) $ is then given by
\begin{equation}
    \sigma ( \vert m_{ee} \vert )_{ \rm max }
    =   \sigma ( \vert m_{ee} \vert )_{ {\rm Im} ( m_3 Q^* P )=0  } 
    \approx  \vert U_{e 3} \vert^2 \sigma ( m_3 ) = 0.0227 \,  \sigma ( m_3 ).
  \label{cco}
\end{equation}
\subsubsection{ (C):
    $   {1 \over m_1}  \vert U_{e 1} \vert^2 + {1 \over m_2}  \vert U_{e 2} \vert^2 
             -  {1 \over m_3}  \vert U_{e 3} \vert^2 =0$
            \hskip0.1cm  (  $ m_3 = 0.0012 \, {\rm eV} $ )   }
In this case, $ Q =0 $ at the point $ ( \beta, \alpha )=( \pi/2, 0 ) $
   in the $ \beta - \alpha $ plane.
Because $ { \partial \vert m_{ee} \vert }/{ \partial m_3 } = 0$ at this point from Eq.(\ref{cbd}),
   one has
\begin{equation}
    \sigma ( \vert m_{ee} \vert )_{ \rm min }
    =   \sigma ( \vert m_{ee} \vert )_{ (\beta, \alpha)=(\pi/2, 0) } 
    =  0.
  \label{ccp}
\end{equation}
The maximum value of $  \sigma ( \vert m_{ee} \vert ) $ in the $ \beta - \alpha $ plane
   becomes
\begin{equation}
    \sigma ( \vert m_{ee} \vert )_{ \rm max }
     =    \sigma ( \vert m_{ee} \vert )_{ (\beta, \alpha)=( 0, 0 ) }
     =   \left\{ 2 \vert U_{e 3} \vert^2   \right \} \sigma ( m_3 ) 
     =   0.0454 \,  \sigma ( m_3 ).
  \label{ccq}
\end{equation}
\subsubsection{ (D):
    $   {1 \over m_1}  \vert U_{e 1} \vert^2 - {1 \over m_2}  \vert U_{e 2} \vert^2 
             -  {1 \over m_3}  \vert U_{e 3} \vert^2 =0$
            \hskip0.1cm  (  $ m_3 = 0.0029 \, {\rm eV} $ )   }
In this case, $ Q =0 $ at the point $ ( \beta, \alpha )=( \pi/2,  \pi/2 ) $.
Since $ { \partial \vert m_{ee} \vert }/{ \partial m_3 } = 0$ at this point from Eq.(\ref{cbd}),
   we have
\begin{equation}
    \sigma ( \vert m_{ee} \vert )_{ \rm min }
    =   \sigma ( \vert m_{ee} \vert )_{ (\beta, \alpha)=(\pi/2, \pi/2) } 
    =  0.
  \label{ccr}
\end{equation}
The maximum value of $  \sigma ( \vert m_{ee} \vert ) $ becomes
\begin{equation}
    \sigma ( \vert m_{ee} \vert )_{ \rm max }
      = \sigma ( \vert m_{ee} \vert )_{ (\beta, \alpha)=( 0, 0 ) }            \nonumber \\                                     
      =   \left\{   2  \, {m_3 \over m_1} \, \vert U_{e 1} \vert^2   \right \}
            \sigma ( m_3 )       \nonumber \\
      =   0.079 \,  \sigma ( m_3 ).
  \label{ccs}
\end{equation}
\subsubsection{ (E):  $ {1 \over m_2 } \vert U_{e 2} \vert^2  \gg
                                     {1 \over m_3 } \vert U_{e 3} \vert^2 $
         and  $   m_3^2
                \mathrel{ \rlap{\raise 0.511ex \hbox{$<$}}{\lower 0.511ex \hbox{$\sim$}}}
                  \bigtriangleup  m_{\rm A}^2  $
       \hskip0.1cm ( $  0.004 \, {\rm eV} \ll  m_3 
                            \mathrel{ \rlap{\raise 0.511ex \hbox{$<$}}{\lower 0.511ex \hbox{$\sim$}}}
                              0.05 \, {\rm eV}   $ )   }

In this case, we obtain the relation,
   $  \vert U_{e 2} \vert^2  \gg  (m_2 / m_3)  \vert U_{e 3} \vert^2 
        >   (m_3 / m_2) \vert U_{e 3} \vert^2  $,
   which allows us to use the approximation,
\begin{eqnarray}
  P  & \approx &  m_1  \vert U_{e 1} \vert^2 +m_2  \vert U_{e 2} \vert^2  e^{2 i \alpha},
                                                           \nonumber \\
  Q  & \approx &  {1 \over m_1 }  \vert U_{e 1} \vert^2 
        +{1 \over m_2}  \vert U_{e 2} \vert^2  e^{2 i \alpha}.
  \label{cct}
\end{eqnarray}
The factor ${ \partial \vert m_{ee} \vert }/{ \partial m_3 }$ is then approximated,
\begin{equation}
    \frac{ \partial \vert m_{ee} \vert }{ \partial m_3 } \,
      \approx  \,
      \frac{m_3  \left\{  \vert U_{e 1} \vert^4 +  \vert U_{e 2} \vert^4
                 + \left( {m_2 \over m_1} +  {m_1 \over m_2} \right)
                  \vert U_{e 1} \vert^2  \vert U_{e 2} \vert^2 \cos 2 \alpha \right\} }
             { ( m_1  \vert U_{e 1} \vert^2 )^2 +  ( m_2  \vert U_{e 2} \vert^2 )^2
                 + 2  ( m_1  \vert U_{e 1} \vert^2 ) ( m_2  \vert U_{e 2} \vert^2 ) \cos 2 \alpha }.
       \label{ccu}
\end{equation}
The factor ${ \partial \vert m_{ee} \vert }/{ \partial m_3 }$ scarcely depends on the
   Majorana phase $\beta$, and is a monotone decreasing function of the Majorana
   phase $\alpha \, ( 0 \leq \alpha \leq \pi / 2 ) $.
The ${ \partial \vert m_{ee} \vert }/{ \partial m_3 }$ takes the maximum value
   $(m_3 / m_1) \vert U_{e 1} \vert^2 + (m_3 / m_2)  \vert U_{e 2} \vert^2
           + O( \vert U_{e 3} \vert^2 )$
   at $ \alpha = 0$, and the minimum value
   $(m_3 / m_1)\vert U_{e 1} \vert^2 - (m_3 / m_2)  \vert U_{e 2} \vert^2
             + O( \vert U_{e 3} \vert^2 )$
   at $ \alpha = \pi/2$.
In other words, $\sigma ( \vert m_{ee} \vert ) $ scarcely depends on $\beta$, and is
   a monotone decreasing function of $\alpha \, ( 0 \leq \alpha \leq \pi / 2 ) $.
The $\sigma ( \vert m_{ee} \vert ) $ takes the maximum value at $ \alpha = 0$,
\begin{eqnarray}
    \sigma ( \vert m_{ee} \vert )_{ \rm max }
    & = &   \sigma ( \vert m_{ee} \vert )_{ \alpha =0 }            \nonumber \\
    &  \approx &   \left\{  {m_3 \over m_1}  \vert U_{e 1} \vert^2 
                      + {m_3 \over m_1}  \vert U_{e 2} \vert^2 \right \}  \sigma ( m_3 )     \nonumber \\                  
    & = &  \left\{  0.977 \,  \frac{m_3}{ \sqrt{ m_3^2 +   \bigtriangleup  m_{\rm A}^2 }   }
                  \right \}  \sigma ( m_3 ),
  \label{ccv}
\end{eqnarray}
and the minimum value at $\alpha=\pi / 2$,
\begin{eqnarray}
    \sigma ( \vert m_{ee} \vert )_{ \rm min }
    & = &   \sigma ( \vert m_{ee} \vert )_{ \alpha =\pi/2 }            \nonumber \\
    &  \approx &   \left\{  {m_3 \over m_1}  \vert U_{e 1} \vert^2 
                      - {m_3 \over m_1}  \vert U_{e 2} \vert^2 \right \}  \sigma ( m_3 )     \nonumber \\
    & = &  \left\{  0.387 \,  \frac{m_3}{ \sqrt{ m_3^2 +   \bigtriangleup  m_{\rm A}^2 }   }
                  \right \}  \sigma ( m_3 ),
  \label{ccw}
\end{eqnarray}
where we have used the approximation $m_1 \approx m_2$,
   and neglected the term of the order $ O( \vert U_{e 3} \vert^2 ) $
   in ${ \partial \vert m_{ee} \vert }/{ \partial m_3 }$.
\subsubsection{ (F): $ m_3^2 \gg { \bigtriangleup  m_{\rm A}^2 } $
                          such that $ m_1 \approx m_2 \approx m_3 $
                          \hskip0.1cm  $( m_3^2  \gg ( 0.05 {\rm eV} )^2 )$   }
Using the relation $ Q \approx  P / m_3^2 $, we have
\begin{equation}
     \frac{ \partial \, \vert m_{ee} \vert }{ \partial \, m_3 } 
     \approx \frac{  \vert m_{ee} \vert }{m_3},
     \label{ccx}
\end{equation}
or
\begin{equation}
   \frac{ \sigma ( \vert m_{ee} \vert ) }{ \vert m_{ee} \vert }
   \approx  \frac{ \sigma ( m_3 ) }{ m_3 }.
  \label{ccy}
\end{equation}
As in the case of $ m_1 \approx m_2 \approx m_3 $ in the normal mass ordering,
   the relative error of the effective mass $  \vert m_{ee} \vert $ has the same magnitude
   as that of the lightest neutrino mass $m_3$, and the following relations hold,
\begin{eqnarray}
    \sigma ( \vert m_{ee} \vert )_{ \rm max }
      & = &  \sigma ( \vert m_{ee} \vert )_{ (\beta, \alpha)=(0,0) } \approx  \sigma ( m_3 ),
                                                                        \nonumber \\
    \sigma ( \vert m_{ee} \vert )_{ \rm min }
      & = &    \sigma ( \vert m_{ee} \vert )_{ (\beta, \alpha)=( \pi/2, \pi/2) }  
                  \approx  0.364 \,  \sigma ( m_3 ).
  \label{ccz}
\end{eqnarray}
\section{ Minimum condition of $ \sigma ( m_0 ) $ for probing
                the Majorana phases. }
What values the Majorana phases $ \beta $, $ \alpha $ take is an interesting subject.
When the values of $ \beta $ and $ \alpha $ are changed, the effective mass
   $ \vert m_{ee} \vert $ varies from $ \vert m_{ee} \vert_{ \rm min } $
   to $ \vert m_{ee} \vert_{ \rm max } $, where we have denoted the maximum value
   of $ \vert m_{ee} \vert $ in the $ \beta-\alpha $ plane as $ \vert m_{ee} \vert_{ \rm max } $
   and the minimum value as $ \vert m_{ee} \vert_{ \rm min } $.
Up to now, only the upper limit of the effective mass $ \vert m_{ee} \vert $ is reported
   in experiments, as well as the upper limit of the lightest neutrino mass is.
Assuming that the lightest neutrino mass $m_0$ is measured $ m_0 \pm \sigma ( m_0 )$,
   we studied in section 2 and 3 how the predicted value of $ \vert m_{ee} \vert $ is given.
The predicted value of $ \vert m_{ee} \vert $,
\begin{equation}
    \vert m_{ee} \vert \pm  n \, \sigma ( \vert m_{ee} \vert ),
        \hskip1cm  (n=1,2,3, \cdots ),
   \label{daa}
\end{equation}
depends on the values of Majorana phases $ \beta $ and $ \alpha $.
The larger the uncertainty $\sigma( m_0 )$ is, the larger becomes the uncertainty
   $  \sigma ( \vert m_{ee} \vert ) $ of the predicted value of $ \vert m_{ee} \vert $.
If the uncertainty $  \sigma ( \vert m_{ee} \vert ) $ is large such that it exceeds the order
   of $ ( \vert m_{ee} \vert_{ \rm max } -  \vert m_{ee} \vert_{ \rm min } ) $,
   one can not obtain the information on the Majorana phases $ \beta $ and $ \alpha $
   through $\vert m_{ee} \vert$.
In this section, we study the minimum condition which the relative error of the lightest
   neutrino mass $m_0$ needs to satisfy, in order to obtain some information on the
   Majorana phases $ \beta $ and $ \alpha $ through the effective mass
    $ \vert m_{ee} \vert $.
    
Let us pay attention to the maximum value $ \sigma ( \vert m_{ee} \vert )_{ \rm max }$
   of $ \sigma ( \vert m_{ee} \vert )$
   in the $ \beta-\alpha $ plane and the point $(\beta, \alpha)$
   at which $ \sigma ( \vert m_{ee} \vert )_{ \rm max }$ is realized.
If the predicted value of $ \vert m_{ee} \vert $ at this point,
   $ \vert m_{ee} \vert \pm n \, \sigma ( \vert m_{ee} \vert )_{ \rm max } $,
   satisfies the following relations simultaneously,
\begin{eqnarray}
    & &  \vert m_{ee} \vert_{\rm max}  
              \le   \vert m_{ee} \vert + n \, \sigma ( \vert m_{ee} \vert )_{ \rm max },
                                                          \\ \nonumber
   & &  \vert m_{ee} \vert - n \, \sigma ( \vert m_{ee} \vert )_{ \rm max }
              \le  \vert m_{ee} \vert_{\rm min},
  \label{dab}
\end{eqnarray}
the information on the Majorana phases $ \beta $ and $ \alpha $ is not obtained
   effectively, because the allowed region of the predicted value of $ \vert m_{ee} \vert $
   is too wide.
As discussed in section 2 and 3, $  \sigma ( \vert m_{ee} \vert ) $ takes its maximum
   value $ \sigma ( \vert m_{ee} \vert )_{ \rm max } $ at the point $  (\beta, \alpha)=(0,0)  $,
   at which the effective mass $ \vert m_{ee} \vert $ takes its maximum value
   $  \vert m_{ee} \vert_{ \rm max }  $.
One therefore obtains from Eq.(77),
\begin{equation}
    \vert m_{ee} \vert_{\rm max} - n \, \sigma ( \vert m_{ee} \vert )_{ \rm max }
    \le  \vert m_{ee} \vert_{\rm min},
  \label{dad}
\end{equation}
or
\begin{equation}
  {1 \over n}  \le   \frac{  \sigma ( \vert m_{ee} \vert )_{ \rm max } }
                           {  \vert m_{ee} \vert_{\rm max} -   \vert m_{ee} \vert_{\rm min}  }.
  \label{dae}
\end{equation}
When $\sigma ( \vert m_{ee} \vert )_{ \rm max }$ has large value such as
   Eq.(\ref{dae}), we can not obtain the information on the Majorana phases
   $ \beta $ and $ \alpha $ effectively because of the wide allowed region of the predicted
   value of $ \vert m_{ee} \vert $.
The maximum value $\sigma ( \vert m_{ee} \vert )_{ \rm max }$ is represented by the
   uncertainty $\sigma ( m_0 )$ of the lightest neutrino mass $m_0$ if one put the
   uncertainties of the oscillation parameters zero.
We can therefore restrict the uncertainty $\sigma ( m_0 )$ of $m_0$ in order to obtain
   the information on the Majorana phases $ \beta $ and $ \alpha $.
In the following subsections, the uncertainty $\sigma ( m_0 )$ is restricted in the normal
   mass ordering case and the inverted mass ordering case, respectively.
\subsection{ In the normal mass ordering case. }
In this subsection dealing with the normal mass ordering case, we consider the lightest
   neutrino mass $m_1$ that satisfies
\begin{equation}
                 m_2  \vert U_{e 2} \vert^2 + m_3  \vert U_{e 3} \vert^2  
              <     m_1 \vert U_{e 1} \vert^2,
  \label{dba}
\end{equation}
or
\begin{equation}
   m_1  >   0.0063 \, {\rm eV}.
  \label{dbb}
\end{equation}
Since the lower limit of $m_1$ is not given by experiments, rather large value of $m_1$
   is interesting for the present.
The effective mass $ \vert m_{ee} \vert $ takes its maximum value at the point
   $  (\beta, \alpha)=(0,0)  $,
\begin{equation}
   \vert m_{ee} \vert_{\rm max}
     =   m_1 \vert U_{e 1} \vert^2 + m_2  \vert U_{e 2} \vert^2 + m_3  \vert U_{e 3} \vert^2,
   \label{dbc}
\end{equation}
while it takes its minimum value at the point
   $ (\beta, \alpha)=( \pi / 2, \pi / 2 )$ \cite{rf:NunTevFun,rf:Mae},
\begin{equation}
   \vert m_{ee} \vert_{\rm min}
     =   m_1 \vert U_{e 1} \vert^2 - m_2  \vert U_{e 2} \vert^2 - m_3  \vert U_{e 3} \vert^2,
   \label{dbd}
\end{equation}
for $m_1$ satisfying Eq.(\ref{dba}).
The width between $ \vert m_{ee} \vert_{\rm max} $ and $ \vert m_{ee} \vert_{\rm min} $
   then becomes
\begin{equation}
   \vert m_{ee} \vert_{\rm max} -   \vert m_{ee} \vert_{\rm min}
      =  2 ( m_2  \vert U_{e 2} \vert^2 + m_3   \vert U_{e 3} \vert^2  ).
  \label{dbe}
\end{equation}
Using Eq.(\ref{dbe}) and Eq.(\ref{bbi}), we get
\begin{equation}
   \frac{  \sigma ( \vert m_{ee} \vert )_{ \rm max } }
          {  \vert m_{ee} \vert_{\rm max} -   \vert m_{ee} \vert_{\rm min}  }  
    =   \frac{ m_1  \left\{  \vert U_{e 1} \vert^2
                                        + {m_1 \over m_2}  \vert U_{e 2} \vert^2 
                                        +  {m_1 \over m_3}  \vert U_{e 3} \vert^2 \right\}             }
                    {  2 ( m_2  \vert U_{e 2} \vert^2 + m_3   \vert U_{e 3} \vert^2  ) }   
     \times \left\{ \frac{ \sigma( m_1 ) }{ m_1 } \right\}.
  \label{dbf}
\end{equation}
In the right-handed side of this equation, the coefficient of $  \sigma( m_1 ) / m_1 $
   is a monotone increasing function of the lightest neutrino mass
   $m_1 \, ( > 0.0065 {\rm eV} ) $.
Let us consider two concrete cases; one is the case where
   $ m_1 \gg \sqrt{ \bigtriangleup  m_{\rm A}^2 } \sim 0.05 {\rm eV} $,
   and the other is the case where
   $ m_1 = \sqrt{ \bigtriangleup  m_{\rm A}^2 } \sim 0.05 {\rm eV} $.
In the first case, $ m_1 \gg \sqrt{ \bigtriangleup  m_{\rm A}^2 } \sim 0.05 {\rm eV} $,
   three neutrino masses are degenerate, $ m_1 \approx m_2 \approx m_3 $.
When $ n=2 $, Eq.(\ref{dae}) becomes
\begin{equation}
   {1 \over 2} \leq 1.57 \, \left\{ \frac{ \sigma( m_1 ) }{ m_1 } \right\},
  \label{dbg}
\end{equation}
which implies that if the relative error $  \sigma( m_1 ) / m_1 $ is larger than
   $ 32 \% $, the relation
   $ \vert m_{ee} \vert_{\rm max} - 2 \, \sigma ( \vert m_{ee} \vert )_{ \rm max }
              \le  \vert m_{ee} \vert_{\rm min} $
   (Eq.(\ref{dab})) holds, and the information on the Majorana phases $ \beta $
   and $ \alpha $ is not obtained effectively.
Thereby, in the case $ m_1 \gg \sqrt{ \bigtriangleup  m_{\rm A}^2 } \sim 0.05 {\rm eV} $,
   the minimum condition so as to
   raise the possibility of obtaining the information on $ \beta $ and $ \alpha $ is that the
   relative error $  \sigma( m_1 ) / m_1 $ of the lightest neutrino mass should be less
   than $ 32 \% $ at least.
In the second case,
   $ m_1 = \sqrt{ \bigtriangleup  m_{\rm A}^2 } \sim 0.05 {\rm eV} $, Eq.(\ref{dae}) with
   $ n=2 $ becomes
\begin{equation}
   {1 \over 2} \leq 1.48 \, \left\{ \frac{ \sigma( m_1 ) }{ m_1 } \right\}.
  \label{dbh}
\end{equation}
Therefore, in the case $ m_1 = \sqrt{ \bigtriangleup  m_{\rm A}^2 } \sim 0.05 {\rm eV} $,
   the minimum condition so as to raise the possibility of obtaining the information on
   $ \beta $ and $ \alpha $ is that the relative error $  \sigma( m_1 ) / m_1 $ should
   be less than $ 34 \% $ at least.
\subsection{ In the inverted mass ordering case. }
The effective mass $ \vert m_{ee} \vert $ takes the maximum value at the point
   $  (\beta, \alpha)=(0,0)  $,
\begin{equation}
   \vert m_{ee} \vert_{\rm max}
     =   m_1 \vert U_{e 1} \vert^2 + m_2  \vert U_{e 2} \vert^2 + m_3  \vert U_{e 3} \vert^2,
   \label{dca}
\end{equation}
while it takes the minimum value at the point $  (\beta, \alpha)=( \pi / 2, \pi / 2 )  $,
\begin{equation}
   \vert m_{ee} \vert_{\rm min}
     =   m_1 \vert U_{e 1} \vert^2 - m_2  \vert U_{e 2} \vert^2 - m_3  \vert U_{e 3} \vert^2,
   \label{dcb}
\end{equation}
and the width becomes
\begin{equation}
   \vert m_{ee} \vert_{\rm max} -   \vert m_{ee} \vert_{\rm min}
      =  2 ( m_2  \vert U_{e 2} \vert^2 + m_3   \vert U_{e 3} \vert^2  ).
  \label{dcc}
\end{equation}
From this equation and Eq.(\ref{cbf}), one has
\begin{equation}
   \frac{  \sigma ( \vert m_{ee} \vert )_{ \rm max } }
          {  \vert m_{ee} \vert_{\rm max} -   \vert m_{ee} \vert_{\rm min}  }  
    =   \frac{ m_3  \left\{  {m_3 \over m_1}  \vert U_{e 1} \vert^2
                                        + {m_3 \over m_2}  \vert U_{e 2} \vert^2 
                                        +  \vert U_{e 3} \vert^2 \right\}             }
                    {  2 ( m_2  \vert U_{e 2} \vert^2 + m_3   \vert U_{e 3} \vert^2  ) }   
     \times \left\{ \frac{ \sigma( m_3 ) }{ m_3 } \right\}.
  \label{dcd}
\end{equation}
In the right-handed side of this equation, the coefficient of $  \sigma( m_3 ) / m_3 $
   is a monotone increasing function of the lightest neutrino mass $m_3$.
Two concrete cases are considered here; one is 
   $ m_3 \gg \sqrt{ \bigtriangleup  m_{\rm A}^2 } \sim 0.05 {\rm eV} $, and the other is
   $ m_3 = \sqrt{ \bigtriangleup  m_{\rm A}^2 } \sim 0.05 {\rm eV} $.
In the first case, three neutrino masses are degenerate and Eq.(\ref{dae}) with
   $ n=2 $ becomes
\begin{equation}
   {1 \over 2} \leq 1.57 \, \left\{ \frac{ \sigma( m_3 ) }{ m_3 } \right\}.
  \label{dce}
\end{equation}
Hence, in the case $ m_3 \gg 0.05 {\rm eV} $, the minimum condition so as to raise
   the possibility of obtaining the information on $ \beta $ and $ \alpha $ is that
   the relative error $  \sigma( m_3 ) / m_3 $ of the lightest neutrino mass $m_3$
   should be less than $ 32 \% $ at least.
The above result with the lightest neutrino mass
   $ m_3 \gg \sqrt{ \bigtriangleup  m_{\rm A}^2 } $ in the inverted mass ordering case
   is the same as that with the lightest neutrino mass
   $ m_1 \gg \sqrt{ \bigtriangleup  m_{\rm A}^2 } $ in the normal mass ordering case.
This is natural because three neutrino masses are degenerate
   $ m_1 \approx m_2 \approx m_3 $ in both cases.
In the second case, $ m_3 = \sqrt{ \bigtriangleup  m_{\rm A}^2 } \sim 0.05 {\rm eV} $,
   Eq.(\ref{dae}) with $ n=2 $ becomes
\begin{equation}
   {1 \over 2} \leq 0.81 \, \left\{ \frac{ \sigma( m_3 ) }{ m_3 } \right\}.
  \label{dcf}
\end{equation}
Then, in the case $ m_3  \sim 0.05 {\rm eV} $, the minimum condition so as to raise
   the possibility of obtaining the information on $ \beta $ and $ \alpha $ is that the
   relative error $  \sigma( m_3 ) / m_3 $ should be less than $ 62 \% $ at least.

We shall make two comments here.
The first comment is concerned with the conditions Eq.(\ref{dcf}) and Eq.(\ref{dbh})
   when the lightest neutrino mass is about $ 0.05 {\rm eV} $.
Eq.(\ref{dcf}) shows that the relative error of the lightest neutrino mass should be less
   than $62 \%$ in the inverted mass ordering case,
   while Eq.(\ref{dbh}) shows that the relative error of the lightest neutrino mass should
   be less than $34 \%$ in the normal mass ordering case.
The contrast between these comes mainly from the fact that the behavior of
   $\sigma ( \vert m_{ee} \vert )_{ \rm max } / \sigma (m_0)$ in the inverted
   mass ordering case is different from that in the normal mass ordering case.
As seen from Fig.1, the value of $\sigma ( \vert m_{ee} \vert )_{ \rm max } / \sigma (m_0)$
   in the inverted mass ordering case is considerably less than the value of that in
   the normal mass ordering case in the region
   $ m_0  \mathrel{ \rlap{\raise 0.511ex \hbox{$<$}}{\lower 0.511ex \hbox{$\sim$}}}
                        0.05 \, {\rm eV}  $.
The second comment is about the minimum condition for the relative error
   $ \sigma( m_0 ) / m_0 $ to raise the possibility of obtaining the information on
   $ \beta $ and $ \alpha $.
In this paper,we have set the errors of the oscillation parameters zero.
When these errors are included in our calculation, the condition required for the
   relative error of the lightest neutrino mass $m_0$ becomes stricter.
Hence, the condition for $\sigma (m_0) / m_0$ obtained in this chapter should be regarded
   as the least necessary condition.
\section{Conclusion}
We studied the influence of the error of the lightest neutrino mass $m_0$ on
   the predicted value of the effective Majorana mass $  \vert m_{ee} \vert  $
   under the assumption that the lightest neutrino mass is measured in experiments.
The case of assuming the normal mass ordering and the case of assuming the
   inverted mass ordering are studied, respectively.
The error of the predicted value of the effective Majorana mass
   $ \sigma ( \vert m_{ee} \vert ) $ is represented by the error of the lightest neutrino
   mass $\sigma( m_0 )$ with the law of propagation of errors provided that one sets the
   errors of the oscillation parameters zero, respectively.
We have studied the error $ \sigma ( \vert m_{ee} \vert ) $ depending on the
   Majorana phases $ \beta $ and $ \alpha $ how it behaves in the $ \beta - \alpha $ plane,
   and how it changes according to the size of the lightest neutrino mass $m_0$.
It is shown that the maximum value of $ \sigma ( \vert m_{ee} \vert ) $ in the
   $ \beta - \alpha $ plane does not exceed the value of $\sigma( m_0 )$ in general,
   and that the minimum value of $ \sigma ( \vert m_{ee} \vert ) $ in the
   $ \beta - \alpha $ plane can be zero when the lightest neutrino mass $m_0$ is small.
An interesting feature is seen when the lightest neutrino mass $m_0$ has large value
   $m_0^2 \gg  \bigtriangleup  m_{\rm A}^2 $ so that three neutrino masses are degenerate.
In this case, the relative error of $  \vert m_{ee} \vert  $ and that of $m_0$ are the same.
We also investigated how the maximum value of $ \sigma ( \vert m_{ee} \vert ) $
   in the $ \beta - \alpha $ plane changes according to the size of the lightest neutrino
   mass $m_0$.
In the normal mass ordering case, the maximum value of $ \sigma ( \vert m_{ee} \vert ) $
   in the $ \beta - \alpha $ plane is almost the same as $ \sigma (m_0) $ when the lightest
   neutrino mass $m_0 = m_1$ is large such as
   $m_1^2  \gg  \bigtriangleup  m_{\rm A}^2 $.
When the value of $m_0 = m_1$ decreases, the maximum value of
   $ \sigma ( \vert m_{ee} \vert ) $ decreases gradually, and it becomes
   $  \vert U_{e 1} \vert^2 \, \sigma (m_0) \sim 0.682 \, \sigma (m_0) $ if $m_1$ is
   approximately zero.
In the inverted mass ordering case, the maximum value of $ \sigma ( \vert m_{ee} \vert ) $
   in the $ \beta - \alpha $ plane is almost the same as $ \sigma (m_0) $ when the lightest
   neutrino mass $m_0 = m_3$ is large such as
   $m_3^2  \gg  \bigtriangleup  m_{\rm A}^2 $.
When the value of $m_0 = m_3$ decreases, however, the maximum value of
   $ \sigma ( \vert m_{ee} \vert ) $ decreases rapidly, and it becomes
   $  \vert U_{e 3} \vert^2 \, \sigma (m_0) \sim 0.0227 \, \sigma (m_0) $ if $m_3$ is
   approximately zero.
   
In Eq.(\ref{bab}) where $ \sigma ( \vert m_{ee} \vert ) $ is represented by $ \sigma (m_0) $
   (the case of $m_0 = m_1$), the factor ${ \partial \vert m_{ee} \vert }/{ \partial m_0 }$
   $(m_0=m_1)$ has been calculated straightforwardly and the result Eq.(\ref{bba})
   contains $ \vert m_{ee} \vert $ in the denominator.
When $ \vert m_{ee} \vert $ becomes very small, therefore, one may guess that
   $ \sigma ( \vert m_{ee} \vert ) $ can take very large value at first sight.
However, it is shown by the analytic calculation that the maximum value of
   $ \sigma ( \vert m_{ee} \vert ) $ in the $ \beta - \alpha $ plane does not exceed
   $ \sigma (m_0) $.
This demonstration owes the complex quantities $P$ and $Q$ which are introduced in
   Eq.(\ref{bbe}) and Eq.(\ref{bbf}).
Utilizing these $P$ and $Q$, we showed that the maximum value of 
   $ \sigma ( \vert m_{ee} \vert ) $ in the $ \beta - \alpha $ plane is given by Eq.(\ref{bbi})
   in the normal mass ordering case, and by Eq.(\ref{cbf})
   in the inverted mass ordering case, respectively.

We have especially paid attention to the influence of the error of the lightest neutrino mass
   $ \sigma (m_0) $ on the uncertainty of the predicted value of the effective
   Majorana mass $ \sigma ( \vert m_{ee} \vert ) $.
If we wish to have the information on the Majorana phases $ \beta $ and $ \alpha $
   through $ \vert m_{ee} \vert $,
   we need to investigate the dependence of $ \vert m_{ee} \vert $ on
   $ \beta $ and $ \alpha $.
Hence, we studied the uncertainty $ \sigma ( \vert m_{ee} \vert ) $ of the predicted value of
   $ \vert m_{ee} \vert $ in detail; how does it behave in the $ \beta - \alpha $ plane,
   how does it depend on the lightest neutrino mass $m_0$, and how large is it.
At the outset, the minimum condition required for the relative error
   $ \sigma (m_0) / m_0$ in order to raise the possibility of obtaining the information on
   $ \beta $ and $ \alpha $ is considered in section 4.
For instance, in the case of $ m_0^2 \gg { \bigtriangleup  m_{\rm A}^2 } $
   ($ m_1 \approx m_2 \approx m_3 $), the relative error of $m_0$ should be less than
   $32 \%$ so that the allowed region of $ \vert m_{ee} \vert $ does not cover the width
   $ ( \vert m_{ee} \vert_{\rm max} -   \vert m_{ee} \vert_{\rm min} ) $ by $ 2 \sigma $.
   
Hereafter, it will be necessary to study minimum condition required for the relative
   error of $m_0$ so as to narrow the allowed region of the predicted value
   $ \vert m_{ee} \vert $.
In the following, we give a simple example in the case
   $ m_0 = m_1 \approx m_2 \approx m_3 $.
If the CP-invariance are maintained, the Majorana phases take the values,
   $ \alpha = n \pi /2, \, \beta = n' \pi /2, \, ( n, n' \in {\bf Z} ) $, and in the restricted
   region $ -\pi/2 < \beta, \alpha \le \pi/2$,
\begin{equation}
      (  \beta, \alpha ) = ( 0, 0), \hskip0.2cm  ( {\pi \over 2}, 0),
                             \hskip0.2cm  (0,  {\pi \over 2} ),  \hskip0.2cm  ( {\pi \over 2},  {\pi \over 2}).
     \label{ea}
\end{equation}
The values of $  \vert m_{ee} \vert  $ at the above four points $ (\beta, \alpha)$ satisfy
\begin{equation}
      \vert m_{ee} \vert_{ ( \beta, \alpha ) = (0, 0) } \, >
      \vert m_{ee} \vert_{ ( \beta, \alpha ) = ( {\pi \over 2} , 0) } \, >
      \vert m_{ee} \vert_{ ( \beta, \alpha ) = ( 0, {\pi \over 2}) } \, >
      \vert m_{ee} \vert_{ ( \beta, \alpha ) = ( {\pi \over 2}, {\pi \over 2}) }.
  \label{eb}
\end{equation}
The allowed region of the predicted value of $\vert m_{ee} \vert_{ ( \beta, \alpha )}$
   is given by
\begin{equation}
       \vert m_{ee} \vert_{ ( \beta, \alpha ) } 
         \pm n \, \sigma(  \vert m_{ee} \vert )_{ ( \beta, \alpha ) },  \hskip1cm ( n=1,2,3, \cdots ).
  \label{ec}
\end{equation}
If the error $ \sigma(m_0) $ of the lightest neutrino mass is large such as 
\begin{equation}
           \vert m_{ee} \vert_{ ( {\pi \over 2} , 0) }
                   - n \,  \sigma(  \vert m_{ee} \vert )_{  ( {\pi \over 2} , 0) }
    \le     \vert m_{ee} \vert_{ ( 0, {\pi \over 2} ) }
      + n \,  \sigma(  \vert m_{ee} \vert )_{  ( 0, {\pi \over 2} ) },
  \label{ed}
\end{equation}
the allowed region of $ \vert m_{ee} \vert_{ ( {\pi \over 2} , 0) } $ and that of
   $  \vert m_{ee} \vert_{ ( 0, {\pi \over 2} ) } $ overlap.
Under this condition, the information on the CP violating Majorana phases
   $ \beta$ and $\alpha $ which lie between $ \vert m_{ee} \vert_{ ( {\pi \over 2} , 0) } $ and
   $  \vert m_{ee} \vert_{ ( 0, {\pi \over 2} ) } $ can not be obtained.
Using the relation
   $ \sigma ( \vert m_{ee} \vert ) = \vert m_{ee} \vert  \{ \sigma ( m_0 ) / m_0 \} $,
   Eq.(\ref{ed}) becomes
\begin{equation}
       \frac{ \sigma( m_0 ) }{ m_0 } 
    \ge
       {1 \over n} \left(  \frac{  \vert U_{e 2} \vert^2 - \vert U_{e 3} \vert^2  }
                                         { \vert U_{e 1} \vert^2 }  \right)
    = 0.20,
  \label{ef}
\end{equation}
where $n=2$ has been adopted.
We therefore see that, in order to obtain the information on $ \beta$ and $\alpha $
   through the $  \vert m_{ee} \vert $ 
   ($ \vert m_{ee} \vert_{ ( {\pi \over 2} , 0) } < \vert m_{ee} \vert 
     <   \vert m_{ee} \vert_{ ( 0, {\pi \over 2} ) } $)
   with CP violating Majorana phases $ \beta, \alpha $, the relative error of $m_0$ should
   be less than $20 \%$ at least.
This is a simple example, and further investigation will be needed.
Lastly, we note that we have set the errors of the measured values of the oscillation
   parameters zero for simplicity.
Although these errors are expected to decrease by future experiments, it is necessary to
   consider the influence of these errors of the oscillation parameters
   on the $  \vert m_{ee} \vert $.
%
%
%
%
%
\newpage
\end{document}